\documentclass[%
showpacs
,12pt,eqsecnum,preprint,secnumarabic,tightenlines
]{revtex4}
\usepackage{acromake}
\usepackage{amssymb}
\def\eg{\emph{e.g.}}                                     
\def\ie{\emph{i.e.}}                                     
\def\etc{\emph{etc}}                                     
\def\wrt{w.r.t.}                                         
\def\d{\mathrm{d}}                                       
\def\e{\mathop{\mathrm{e}}\nolimits}                     
\def\xmin{x_\mathrm{min}}                                
\def\anti#1{\mathpalette{\@anti}{#1}#1}
\def\@anti#1#2{\sbox0{$#1#2$}
  \makebox[0pt][l]{$#1\kern.30\ht0\overline{\kern-.35\ht0\phantom{\bar#2}}$}}
\def\unitop{{\hbox to0.25em{\hrulefill\kern-0.25em       
  \vrule height1.50ex width0.05em\kern0.05em
  \vrule height1.65ex width0.05em\kern-0.25em\kern-0.04em
  \raise0.74ex\hbox{\char'40}}\kern0.25em}}
\def\argleft{\kern-0.25em\left}                           
\makeatletter
  \def\OneTwoCol#1#2{\twocolumn@sw{#2}{#1}}              
\makeatother
\acromake {LLA}  {LLA}  {leading logarithmic approximation}
\acromake {NLLA} {NLLA} {next-to-leading-logarithmic approximation}
\acromake {NS}   {NS}   {non-singlet}
\begin{document}
\title{%
  A Matrix Approach to Numerical Solution \\
  of the DGLAP Evolution Equations}
\author{Philip G. Ratcliffe}
\email{pgr@fis.unico.it}
\affiliation{%
  Dipartimento di Scienze CC.FF.MM., \\
  Universit{\`a} degli Studi dell'Insubria---sede di Como \\
  via Valleggio 11, 22100 Como, Italy \\
  and \\
  Istituto Nazionale di Fisica Nucleare---sezione di Milano \\
  via Celoria 15, 20133 Milano, Italy}
\date{December 2000}
\begin{abstract}
  A matrix-based approach to numerical integration of the DGLAP evolution
  equations is presented. The method arises naturally on discretisation of
  the Bjorken $x$ variable, a necessary procedure for numerical integration.
  Owing to peculiar properties of the matrices involved, the resulting
  equations take on a particularly simple form and may be solved in closed
  analytical form in the variable $t=\ln(\alpha_0/\alpha)$. Such an approach
  affords parametrisation via data $x$ bins, rather than fixed functional
  forms. Thus, with the aid of the full correlation matrix, appraisal of the
  behaviour in different $x$ regions is rendered more transparent and free of
  pollution from unphysical cross-correlations inherent to functional
  parametrisations. Computationally, the entire programme results in greater
  speed and stability; the matrix representation developed is extremely
  compact. Moreover, since the parameter dependence is linear, fitting is very
  stable and may be performed analytically in a single pass over the data
  values.
\end{abstract}
\maketitle
\section{Introduction}

Since the pioneering work of Duke and Owens \cite{Duke:1984x1} in the early
eighties there has been an enormous investment in the phenomenological study
of parton densities. As is well-known, the quark and gluon distributions
inside hadrons acquire a scale ($Q^2$) dependence via higher-order PQCD
corrections. Account of such dependence is necessary, both for correct
analysis of experimental data and for various questions of theoretical
importance. The evolution of the parton densities with $Q^2$ is governed by
the DGLAP equations \cite{Gribov:1972ri, Altarelli:1977zs,
Dokshitzer:1977sg}, the integro-differential nature of which hampers their
numerical solution. In the literature there exist numerous techniques: such
as, the so-called brute-force method \cite{Hirai:1998gb}, the use of Laguerre
polynomials \cite{Kumano:1992x1, Coriano:1998wj} and solution in Mellin
moment space with subsequent inversion. Short-comings common to almost all
are the computer time required and decreasing accuracy for $x\to0$.

The precise motivation for developing this approach was an attempt to avoid
the strong interplay between parameters (and $x$ regions) to which standard
approaches are prone. For example, typically, in any standard approach the
overall normalisation is a function of \emph{all} available parameters and
thus, in particular, the low-$x$ asymptotic power is prisoner to any
fluctuations that may occur in the high-$x$ region. As will emerge, the
natural solution to this problem, \ie, to use function values themselves
(binned in $x$ and $Q^2$) as the parameters, leads to a gross simplification
of the integro-differential equations involved. Indeed, suitable
discretisation of the $x$ variable immediately allows \emph{exact}
integration of the differential equations in $Q^2$ and, moreover, most
calculations may be performed \emph{prior} to actual evolution or data
fitting. A further enormous simplification of the equations is obtained by a
suitable choice of the $x$ bins---the matrices governing evolution are then
``banded'' lower triangular, with two crucial effects: they commute (and may be
expressed in terms of sums of the unit matrix and nil-potent matrices);
moreover, multiplication of such matrices is an order $n^2$ operation (as
opposed to the usual $n^3$) and they may be stored in very compact form, \ie,
storage requirements are order $n$ (as opposed to the usual $n^2$).

Before proceeding it should be pointed out that an approach related to that
proposed here has been presented by Santorelli and Scrimieri
\cite{Santorelli:1998yt}. Rather than comment immediately on the similarities
and differences, it is probably more convenient to highlight such in the
following, as and when appropriate.

This paper is structured as follows: the following section contains the basic
derivation of the approach to evolution while the subsequent covers its
extension to singlet densities and higher-orders; in section~\ref{sec:numres}
numerical results are presented and section~\ref{sec:datfit} describes the
application to data fitting. Finally, some conclusions and prospects for the
technique are presented. The appendices contain a brief demonstration of the
special properties of the matrices involved and details of kernel integration.
\section{Outline of the Method}

In the following the general form of the evolution equations is first
presented and simplifying substitutions of variables are then applied.
Finally, the matrix equation is derived and solved formally.
\subsection{The Evolution Equations}

In the simplest case, \NS to \LLA, the evolution equations for parton
densities take on the form:
\begin{equation}
  \frac{\d f(x,Q^2)}{\d\ln Q^2}
  =
  \frac{\alpha_s(Q^2)}{2\pi}
  \int_x^1 \frac{\d y}{y}\, P\argleft(\frac{x}{y}\right) \, f(y,Q^2) ,
  \label{eq:evol}
\end{equation}
where $f(x,Q^2)$ represents some \NS parton density and $P$ the corresponding
splitting function. The convolution form does allow exact $\ln{Q^2}$
integration if transformed to Mellin moment space. However, this requires
knowledge of the function over the \emph{entire} $x$ region. Thus, if data
analysis is the main aim, it is desirable to remain in $x$ space.
\subsection{Simplifying Substitution of Variables}

Let us start by applying two simplifying variable transformations. It is
convenient to substitute the scale variable $Q^2$ with the variable $t$:
\begin{equation}
  t = \ln \frac{\alpha(Q_0^2)}{\alpha(Q^2)},
  \label{eq:t-def}
\end{equation}
where $Q_0^2$ is some starting scale for evolution (typically a very few
GeV$^2$). The second substitution is for the Bjorken variables $x$ and $y$:
\begin{equation}
  u = \ln \frac1x, \qquad v = \ln \frac1y.
  \label{eq:u-def}
\end{equation}
With these two substitutions, eq.~(\ref{eq:evol}) simplifies to
\begin{equation}
  \frac{\d f(u,t)}{\d t}
  =
  \int_0^u \d v \, P(u-v) \, f(v,t),
  \label{eq:evol-subs}
\end{equation}
where we have taken into account that to \LLA $\alpha_s$ is proportional to
$1/\ln{Q^2}$ and various coefficients have been absorbed into the definition
of $P$.
\subsection{The Matrix Equation and a First Solution}

To examine the numerical approach, consider performing the right-hand side
integral via a na{\"\i}ve trapezoidal rule over subintervals of size $h$:
\begin{equation}
  \frac{\d f_m(t)}{\d t}
  =
  \sum_{k=1}^{m}  \, P_{mk} \, f_k(t),
  \label{eq:evol-mat}
\end{equation}
where the following rather obvious definitions have been made:
\begin{equation}
  u_k = k\,h,
  \OneTwoCol{\quad}{\;}
  f_k(t) = f(u_k,t),
  \OneTwoCol{\quad}{\;}
  P_{mk} = h \, P(u_m-u_k),
  \label{eq:vec-def}
\end{equation}
the typical vanishing of $f(x)$ at $x=1$ has been exploited and a factor one
half in the last term of the series has been omitted. By noting that the sum
in eq.~(\ref{eq:evol-mat}) runs only up to $m$, one sees that the matrix
$P_{mk}$ is lower triangular. Thus, finally, writing the equation in matrix
form, one obtains
\begin{equation}
  \dot\mathbf{f}(t)
  =
  P \, \mathbf{f}(t),
  \label{eq:evol-mat-full}
\end{equation}
where the dot indicates a derivative \wrt\ $t$.

One would now, at least in principle, need only diagonalise $P$, via a matrix
$D$ say: left multiplication by $D^{-1}$ would then result in
\begin{equation}
  D^{-1} \, \dot\mathbf{f}(t)
  =
  D^{-1} P \, D \, D^{-1} \, \mathbf{f}(t).
  \label{eq:evol-diag}
\end{equation}
Defining $\mathbf{\tilde{f}}=D^{-1}\mathbf{f}$ and the diagonalised matrix
$P_D=D^{-1}PD$, the final simple form would therefore be
\begin{equation}
  \dot\mathbf{\tilde{f}}(t)
  =
  P_D \, \mathbf{\tilde{f}}(t).
  \label{eq:evol-final}
\end{equation}
The \emph{exact} solution could thus be written down directly:
\begin{equation}
  \tilde{f}_m(t)
  =
  \e^{\gamma_m t} \, \tilde{f}_m(0),
  \label{eq:evol-sol}
\end{equation}
where the $\gamma_m$ would be just the eigenvalues of the matrix $P$.
Transforming back to the original basis, one would then have
\begin{equation}
  f_m(t)
  =
  \sum_k \e^{\gamma_k t} \, d_{mk} \, \tilde{f}_k(0).
  \label{eq:sol-final}
\end{equation}
This would then be an \emph{exact} solution in $t$ of the differential
equation, \emph{i.e.}, only the $x$ variable having been discretised and
treated numerically.

It turns out that the eigenvalues are very close to one another and so
diagonalisation is very nearly singular. A possible approach might be to
avoid degeneracy by suitable choice of binning and also to work in extended
precision on the computer; however, as we shall now show, this apparent
obstacle may be turned in to a virtue owing to the nature of the matrices
generated.
\subsection{A Better Solution to the Matrix Equation}

The choice of equally spaced bins in $u$ leads, in fact, to total degeneracy.
This impedes matrix diagonalisation and thus the solution must be obtained
differently. However, in compensation, a natural simplification occurs: the
elements $P_{km}$ only depend on the differences $k-m$ (see
appendix~\ref{app:proof} for a proof). Thus, first of all, such matrices may
be stored numerically as vectors, there being only $n$ independent elements.
More specifically, we may define
\begin{equation}
  P_{km}
  =
  p_{k-m},
  \label{eq:evol-mat-vect}
\end{equation}
where $p_k$ is just a vector of length $n$. Moreover, matrix multiplication,
inversion \etc.\ only require O($n^2$) floating-point operations as compared
with the O($n^3$) for general $n{\times}n$ matrix manipulation. The banded triangular
nature of the matrix provides a further remarkable simplification: the set of
all such matrices is abelian. This means that the \NS equation may effectively
be solved algebraically. Thus, $P$ may be treated as a $c$-number in
eq.~(\ref{eq:evol-mat-full}); the solution is then trivial:
\begin{equation}
  \mathbf{f}(t)
  =
  \e^{P\,t} \, \mathbf{f}(0).
  \label{eq:evol-mat-soln}
\end{equation}

There is a further useful property of the matrices appearing here: if the
diagonal elements of such a matrix are zero then the matrix is nilpotent;
\ie, if $A_{ij}=a_{i-j}$ is such an $n{\times}n$ matrix with $a_0=0$ then $A^n=0$.
Separating $P$ according to $P=p_0\,\unitop+\anti{P}$, commutation of the two
terms being trivial, we may then write
\begin{equation}
  \mathbf{f}(t)
  =
  \e^{p_0t} \, \e^{\anti{P}\,t} \, \mathbf{f}(0),
  \label{eq:evol-soln-fact}
\end{equation}
where, owing to the nilpotency of $\anti{P}$, the factor $\exp[\anti{P}\,t]$
is now given by a \emph{finite} sum of power terms. It is easy to see that
its calculation indeed only requires O$(n^3)$ multiplications and is thus
equivalent to a single normal matrix product. Moreover, \emph{any} function
admitting a power-series expansion (\eg, $\sin$, $\cos$, $\log$, square-root,
\etc.) is evaluated similarly. The $n$ terms of any such sum, of course, need
only be calculated once and then stored (as an $n{\times}n$ matrix), providing for
very rapid evaluation for any $t$.
\section{Extension to Singlet and Higher-Order}

In all modern analyses the extension to both singlet and higher logarithmic
accuracy is an absolute necessity. Thus, we now turn first to the extension
to the singlet case and secondly to higher-order QCD corrections.
\subsection{The Singlet Equations}

In the singlet case the equations become a $2{\times}2$ system:
\begin{eqnarray}
  \dot\mathbf{\Sigma}(t)
  &=&
  P_{\Sigma\Sigma} \, \mathbf{\Sigma}(t) + P_{\Sigma g} \, \mathbf{g}(t),
\nonumber
\\
  \dot\mathbf{g}(t)
  &=&
  P_{g\Sigma} \, \mathbf{\Sigma}(t) + P_{gg} \, \mathbf{g}(t),
  \label{eq:evol-mat-sing}
\end{eqnarray}
where the singlet-quark density is $\Sigma=\sum_f{q_f(x)}$ and the $P_{ab}$
are individually of banded triangular form. If we rewrite
eq.~(\ref{eq:evol-mat-sing}) as the outer product of $2{\times}2$ matrices and the
banded triangular kernels, $P_{ab}$,
\begin{equation}
  \dot\mathbf{F}(t)
  =
  \mathcal{P} \, \mathbf{F}(t),
  \label{eq:evol-2x2-mat}
\end{equation}
one immediately sees that the form of the original solution still holds:
namely,
\begin{equation}
  \mathbf{F}(t)
  =
  \exp[\mathcal{P}\,t] \, \mathbf{F}(0).
  \label{eq:evol-2x2-soln}
\end{equation}

Of course, the matrix $\mathcal{P}$, not being of the banded triangular type,
does not permit direct evaluation of the exponential via a finite polynomial.
However, the $2{\times}2$ matrix elements \emph{are} commuting matrices which may
still then be treated as $c$-numbers; the solution is thus trivially obtained
by diagonalisation of the $2{\times}2$ system:
\begin{equation}
  \mathbf{F}(t)
  =
  D \, \exp[\mathcal{P}_D\,t] \, D^{-1} \, \mathbf{F}(0),
  \label{eq:evol-2x2-diag}
\end{equation}
where $\mathcal{P}_D=D^{-1}\mathcal{P}D$ is diagonal in the $2{\times}2$ space. Note
also that the transformation matrix, $D$, may be chosen such that $D^{-1}=D$
(and is real). The upper-left and lower-right diagonal blocks may now be
treated separately along the lines of the \NS case. Note that, for the DGLAP
kernels calculated in QCD, the matrix $\mathcal{P}$ is not singular; in the
case that it were, the solution would in fact be trivial.

Alternatively, we may exploit the simple nature of $2{\times}2$ matrices. Let us use
a basis of Pauli matrices:
\begin{equation}
 \tau_3 = \left( \begin{array}{rr} 1 & 0 \\ 0 &-1 \end{array}\right),
  \OneTwoCol{\quad}{\;}
 \tau_+ = \left( \begin{array}{rr} 0 & 1 \\ 0 & 0 \end{array}\right),
  \OneTwoCol{\quad}{\;}
 \tau_- = \left( \begin{array}{rr} 0 & 0 \\ 1 & 0 \end{array}\right).
\end{equation}
Writing
$\mathcal{P}=P_0\unitop+P_3\tau_3+P_+\tau_++P_-\tau_-=P_0\unitop+\mathbf{P}$,
where the unit matrix and $\mathbf{P}$ now lie in the $2{\times}2$ space and the
$P_a$ lie in $x$-bin space, we obtain
\begin{equation}
  \mathbf{P}^2
  =
  (P_3^2 + P_+P_-) \, \unitop
  \equiv
  \tilde{P}^2 \, \unitop,
\end{equation}
and thus
\begin{equation}
  \e^{\mathcal{P}\,t}
  =
  \e^{P_0\,t} \,
  \left[
    \cosh(\tilde{P}\,t) \, \unitop
    +
    \sinh(\tilde{P}\,t) \, \mathbf{P}/\tilde{P}
  \right].
  \label{eq:evol-2x2-xpnd}
\end{equation}
Note that although $P_0$ and $\tilde{P}$ remain matrices in $x$-bin space
they are also still banded lower triangular and thus the hyperbolic functions
and square-root may again be evaluated as finite polynomials. We note in
passing that, while both forms are stable, the evaluation of
eq.~(\ref{eq:evol-2x2-diag}) requires slightly less computing time than does
(\ref{eq:evol-2x2-xpnd}).
\subsection{Higher Logarithmic Accuracy}
\label{sec:NLLA}

Extending to higher logarithmic accuracy presents no obstacle in the \NS
case, indeed, it is trivial to all orders in $\alpha_s$. The evolution
equation now becomes
\begin{equation}
  \dot\mathbf{q}(t)
  =
  \sum_{n=0}^N \left(\alpha(Q^2)\right)^n \, P^{(n)} \, \mathbf{q}(t),
  \label{eq:evol-ns-ho}
\end{equation}
where the sum runs up to $N$, the order of the approximation. Since the
matrices $P^{(n)}$ are still all of the banded triangular type (\ie,
commuting), an exponential solution of the form of
eq.~(\ref{eq:evol-mat-soln}) is still valid for any $N$:
\begin{equation}
  \mathbf{q}(t)
  =
  \e^{P^{(0)}\,t} \,
  \e^{P^{(1)}\,(1-\e^{-t})} \,
  \e^{P^{(2)}\,\frac12(1-\e^{-2t})} \,
  \cdots \, \mathbf{q}(0).
  \label{eq:evol-soln-ho}
\end{equation}
From the above form, it is also clear that the corrections actually factorise
order-by-order and thus it is even possible to examine their effects
separately and in parallel in a single calculation.

Turning next to the singlet case, we encounter the first and only serious
complication: in
\begin{equation}
  \dot\mathbf{F}(t)
  =
  \sum_{n=0}^N \mathcal{P}^{(n)} \, \e^{-nt} \, \mathbf{F}(t),
  \label{eq:evol-2x2-ho}
\end{equation}
the $2{\times}2$ matrices, $\mathcal{P}^{(n)}$, do \emph{not} commute and thus na{\"\i}ve
exponentiation cannot furnish the required solution. Let us examine the system
in the \NLLA, which we rewrite as follows:
\begin{equation}
  \dot\mathbf{F}(t)
  =
  \left[ \mathcal{A} + \mathcal{B} \, \e^{-t} \right] \, \mathbf{F}(t).
  \label{eq:evol-2x2-nlla}
\end{equation}
If $D$ is the matrix that diagonalises $\mathcal{A}$ at the $2{\times}2$ level, with
the diagonal $\mathcal{A}'_D=D^{-1}\mathcal{A}D$ and
$\mathcal{B}'=D^{-1}\mathcal{B}D$, then defining the vector $\mathbf{G}(t)$
via
\begin{equation}
  \mathbf{F}(t)
  \equiv
  D \, \exp
  \left[
    \mathcal{A}'_D t + \mathcal{B}'_D (1-\e^{-t})
  \right] \, \mathbf{G}(t)
  \label{eq:evol-2x2-sub}
\end{equation}
where $\mathcal{B}'_D$ is the diagonal part of $\mathcal{B}'$ and transforming
to the variable $s\equiv\e^{-t}$ we obtain
\begin{eqnarray}
  \frac{\d\mathbf{G}(s)}{\d s}
  &=& - \exp
  \{ 2
    \left[
      \mathcal{A}'_3 \ln s + \mathcal{B}'_3 (s-1)
    \right]
    \sigma_3
  \}
\OneTwoCol{\,}{\nonumber\\&&\quad\null\times}
  [\mathcal{B}'_+\sigma_+ + \mathcal{B}'_-\sigma_-]
  \, \mathbf{G}(s).
  \label{eq:evol-2x2-red}
\end{eqnarray}

Defining upper and lower elements of $\mathbf{G}$ by $G_{\pm}$, the coupled
first-order equations transform into a pair of decoupled second-order
equations:
\begin{equation}
  \frac{\d^2 G_{\pm}(s)}{\d s^2}
  = {\pm} 2
  \left(
    \frac{\mathcal{A}'_3}{s} + \mathcal{B}'_3
  \right)
  \frac{\d G_{\pm}(s)}{\d s}
  +
  \mathcal{B}'_+ \mathcal{B}'_- \, G_{\pm}(s).
  \label{eq:evol-2x2-2nd}
\end{equation}
The solution may then be obtained in a relatively straight-forward manner via
power-series substitution. However, while the banded triangular nature of the
matrices certainly facilitates evaluation, the series is infinite here and
thus must be truncated. Fortunately, the compact form of the matrices permits
cheap computer storage of a large number of terms.

We note here that the natural expansion is a power series in $s$ and not $t$.
Indeed, while for small $t$ the two are essentially equivalent, for large
values (where $s\to0$) the expansion in $t$ contains only a part of the
series expressed in terms of $s$. In the following section we shall comment
further on this point with regard to the approach of \cite{Santorelli:1998yt}.
\section{Numerical Results on Evolution}
\label{sec:numres}

We have thoroughly tested the approach to \LLA for the both the \NS and full
singlet evolution equations: the performance is highly satisfactory both in
terms of precision and computing time. For the purposes of testing we have
used the following representative set of input distributions at
$Q_0^2=4\,$GeV$^2$:
\begin{eqnarray}
  q_\mathrm{NS}(x,Q_0^2)
  &=&
  A_\mathrm{NS} \, x^{-0.5} \, (1-x)^3,
\nonumber
\\
  \Sigma(x,Q_0^2)
  &=&
  A_\mathrm{S} \, x^{-1} \, (1-x)^3,
\\
  g(x,Q_0^2)
  &=&
  A_g\, x^{-1} \, (1-x)^{6.5}.
\nonumber
\end{eqnarray}
The normalisations are fixed by various sum rules, which also provide means
of cross-checking the accuracy of the method.

As a first examination, let us consider \NS evolution using just 20 points
per decade (\ie, there are 20 bins between any $x$ and $x/10$) and a simple
two-point interpolation to evolve over a range of $t=0$ to 2; this
corresponds to an astronomical final $Q^2=2.3\times10^{12}\,$GeV$^2$ (taking
$Q_0^2=4\,$GeV$^2$ and $\lambda_\mathrm{QCD}=0.240\,$GeV for three active
flavours) but we feel that this is important to test stability of the
algorithm (we shall make further comments later). Under these extreme
conditions (using double-precision arithmetic) the first moment of the
valence quark distribution remains constant to one part in $10^5$ while the
second moments (the momentum sum rule) agree with the analytical evaluation
to 1\% (valence and singlet quark) and 2\% (gluon). Note, however, that to
obtain $10^{-5}$ precision in the first-moment integral it is necessary (for
this number of bins) to cover the region $x\in(10^{-10},1)$ and to employ
higher-order interpolation formul{\ae} for the moment integral itself. The
precision degreases rapidly with increasing moment (owing to heavier sampling
of the large-$x$ region): it is down to the 10\% level by the sixth (fourth)
moment for the valence and singlet quark (gluon).

A significant improvement is obtained by also using higher-order
interpolation formul{\ae} to calculate the matrices. Note that, in order to
maintain the matrix properties, it is necessary to interpolate in each $x$ bin
using only function values at the bin boundaries and \emph{higher} values of
$x$. This means that fake zero values must be introduced for a very few $x$
values larger than one (this essentially cancels the gain for the last bins).
Eight-point interpolation leads to a gain in precision of around three orders
of magnitude for the second moments of all the distributions.

Clearly, a substantial gain may also be made by increasing the density of the
$x$ points, this also permits a check of precision as a function of $x$;
moments are only sensitive to a very limited range of $x$, around the peak of
$x^{n-1}q(x)$. Doubling the density of points to 40 per decade results in an
improvement in precision of a factor 3 (50) for the two-point (8-point)
interpolation. Doubling again to 80 points per decade (\ie, only 400 in the
range $x\in(10^{-5},1)$, the precision for 8-point interpolation is better
than one part in $10^4$ for \emph{all} moments up to $n=12$ and better than
one in $10^6$ for the momentum sum rule, the gain for the higher-order
interpolation is still overwhelming. As a final step we double once more to
160 points per decade, but find no further significant improvement.

Thus, we choose to compare results with a benchmark of 80 points per decade
using 8-point interpolation. For evolution with 20 points per decade and
8-point interpolation, the precision from $x=0.25$ down to $x=10^{-10}$ is
better than 5 parts in $10^6$, for the valence density, still 2\% for the
highest value available for this binning ($x=0.8$) but already 2.4 in $10^4$
for $x=0.5$. The singlet-quark (gluon) density has a roughly factor 20 (30)
poorer accuracy except for the higher $x$ values where the singlet-quark
precision is comparable to that of the valence. As a final test, we compare
evolution with 10 points. Even here performance is excellent: the valence
precision is better than 1 in $10^4$ for $x\le0.2$ while that of the singlet
quark and gluon is better than one per mille over most of the range. Note
that if better precision in the high $x$ region is necessary, then a separate
(parallel) evaluation may be made with a greater concentration of points over
a limited region ($x\ge0.2$ say).

With regard to the large value of $t$ chosen for testing, we now comment on
the method proposed in \cite{Santorelli:1998yt}: there the truncation of the
power series in $t$ at no more than 12 terms is justified on the basis of
rapid convergence. Direct examination of the \emph{full} finite expansion
calculated here suggests this might be illusory: some coefficients do not
become truly small until much later in the series and thus it may be that
convergence is only achieved in \cite{Santorelli:1998yt} thanks to the not
large values of $t$ considered there. We have studied the problem only
superficially, but find that truncation at small order of the series obtained
here renders the procedure unstable and, moreover, the onset of instability
is very sudden with increasing $t$. We should point out, however, the authors
of \cite{Santorelli:1998yt} claim to see no evidence for such behaviour in
their studies \cite{Santorelli:0000pc}.

It might be held that the typical size of $t$ corresponding to physically
accessible energy scales is not normally larger than about $0.3$, as
considered in \cite{Santorelli:1998yt}. On the other hand, if one wishes to
study the dynamical generation of parton distributions starting at some very
small scale (see, for example, \cite{Gluck:1990x2}), then the range of $t$
may be considerably larger. Finally, as commented above, a more suitable
variable would be $s$; in this case the starting point being $s=1$ requires
expansion in the variable $s-1$, which is still always a small variable
($0<s-1\le1$).

We have not yet implemented higher-order corrections; however, in the \NS
case the algorithm will suffer no change, indeed, examination of
eqs.~(\ref{eq:evol-2x2-sub}, \ref{eq:evol-2x2-red}) reveals that higher-order
corrections are implemented as further multiplicative (smaller) corrections.
A possible strategy in the singlet case might be to exponentiate the leading
corrections (possibly together with all diagonal pieces) and then apply the
approach of \cite{Santorelli:1998yt} to the residual equations. Alternatively,
as shown in Section~\ref{sec:NLLA}, after all possible simple factorisation
the remaining differential equation is amenable to power-series solution.
\section{Data Fitting}
\label{sec:datfit}

It is now relatively straight-forward to perform data fitting. The $f_m(0)$ in
eq.~(\ref{eq:evol-soln-fact}) may be taken as the input parameters to fit,
directly providing the parton density at some starting $Q_0^2$. The exact
nature of the solution in $t$ means that to fit a data point at some given
$t$ it is possible to obtain the necessary $f_m$ at precisely that value and
then interpolate in $x$ to the required $x$ value.  Thus, since the matrix
$P$ is calculated once-and-for-all at the start of the fit (and only once
even for many fits), the number of operations necessary to evaluate the fit
function for a given set of values of $x$ and $Q^2$ is drastically reduced
with respect to \emph{all} standard approaches. The number of floating-point
operations may be reduced further by pre-calculating the evolution operator,
$\e^{p_0t}\,\e^{\anti{P}\,t}$, for a set of $t$ values and then interpolating
both in $x$ and $t$. Note that the compact matrix storage requires just a
single vector of length $n$ for each value of $t$. Note too that evolution
may be performed using a larger number of $x$ bins than are used for
parametrisation.

In a fit, since each evolved value, $f(x,t)$, depends only linearly on the
input starting values, $f_m(0)=f(x_m,0)$, $\chi^2$ is quadratic in the
parameters, precisely the $f_m(0)$. Thus, minimisation and calculation of the
full error matrix can, in principle, be performed with a \emph{single} pass
over the data points; this affords considerable saving in computer time and
storage space. Indeed, data values need never be stored, the only memory
requirements being for the evolution matrices (in vector form) and parameter
vectors, together with the resulting covariance matrix.

Moreover, fixing the overall normalisation (\ie, imposing sum rules) at this
stage is entirely superfluous and so has no influence, while the fit will
naturally return a full covariance matrix for the parameters $f_m(0)$, from
which it is then possible to deduce whatever is of interest as to asymptotic
behaviour, satisfaction of sum rules, \emph{etc.}, in an entirely independent
manner.

So far we have only tested a \NS code by generating fake data (with nominal
but realistic errors) at some fixed large $Q^2$. The data generated were not
smeared statistically so that any resulting non-zero $\chi^2$ would be due
purely to inaccuracy in the evolution (based on a limited number of
parameters). As an example we have used 200 data points generated over the
interval $x\in[10^{-5},1]$ and for the same extreme $Q^2$ as above, with 5
equally spaced parameter values per decade (note, however, that there is no
restriction on the parameter spacing) and evolution using 20 points per
decade. We find a total resulting $\chi^2=0.6$ (\ie, 0.004/DoF), the highest
$x$ parameter (corresponding to $x=0.63$) is returned to within 8\% of its
true value while for all $x<0.1$ the precision is better than 2 parts per
mille. The entire fitting procedure (including data reading) takes around one
second on a Pentium-driven laptop. As a final test, we generate an accurate
set of 400 data points over four $Q^2$ slices (equally spaced in $t$) in the
same $x$ range and fit using the same number of parameters and evolution
points as above. The results are very similar to the previous.
\section{Final Comments}

There remain certain technical issues to address, in particular, the matrix
manipulations involved: at any point instabilities may, in principle, arise.
However, this is unlikely; for $h$ small the matrix solution will always be
equivalent to other methods (and, in particular, it must be precisely
equivalent to the so-called brute-force method) and will therefore have a
well-defined and unique solution. Moreover, much of the procedure may be
performed to essentially arbitrary precision beforehand, the resulting
matrices then being used as fixed input to the evolution/fitting procedure.

A further technical issue concerns the discretisation in $x$: the number of
points will always be small, compared with what might normally be adopted for
accurate numerical integration. However, the real data do actually lie in a
finite number of finite-size bins and thus discretisation is in any case
\emph{effectively} forced even in methods where some functional form is fit
to the binned data.

Finally, as discussed, while the extension to \NLLA in the \NS case proves
entirely trivial and to singlet in \LLA is relatively straight-forward (it
becomes a $2\times2$ system of equations), a full extension to the singlet
case in \NLLA proves less direct as the $2\times2$ system is non-commuting.
However, the expansion that is to be truncated should be well-behaved insofar
as the parameter is $\alpha_s$ itself.
\section{Note Added to Completed Manuscript}

On completion of this manuscript the author became aware of unpublished work
by Pascaud and Zomer \cite{Pascaud:1994vx, Pascaud:1996ci}, which contains
many of the basic ideas presented here. There are however certain comments we
should like to make. Firstly, the discussion of the treatment of the high-$x$
region is unnecessarily complicated; it is sufficient to simply choose a
second $x$ grid starting at $x\sim0.2$ with $n\gtrsim100$ and not 1000 via the
complicated formula suggested in \cite{Pascaud:1996ci}. Moreover, the
interpolation adopted there is linear and no comment is made on the
possibility of choosing higher-order polynomials, which, as noted here, do
improve the precision considerably. Finally, and this is the most serious
criticism: as discussed in the present work, a solution of the form of
eq.~(9) in \cite{Pascaud:1996ci} \emph{cannot} be trivially extended to the
$2\times2$ set of singlet equations; the \LLA and \NLLA matrices \emph{do not
commute}, thus direct exponentiation is not possible. In other words, the
expression for $\e^{\mathcal{A}_k}$, despite having the simplified form given
in the cited papers, does \emph{not} correspond to the correct evolution
operator for the singlet densities. Indeed, the authors present only limited
cross-checking of their results using the analytically calculable moments of
the parton distributions, relying purely on the \emph{assumed} accuracy of the
largest-$n$ calculation.
\appendix
\section{The DGLAP Kernels in Numerical Form}
\label{app:kernels}

The kernel matrices may be calculated by hand with no great difficulty. For
simplicity, only the \NS case will be treated here (the singlet case is more
involved numerically but not conceptually). Thus, the equation to solve
implies a convolution integral of some \NS parton distribution, which is
taken to be of the form\footnote{The extra factor $x$ in $xq(x)$, as the
function to evolve, aids in taming small-$x$ singular behaviour typical of
the singlet distributions.} $f(x)\equiv{x}q(x)$, with the following kernel:
\begin{equation}
  P(x)
  \propto
  x \left[ \frac{(1+x^2)}{(1-x)_+} + \frac32 \delta(x-1) \right],
\end{equation}
where the `$+$' regularisation is defined by
\begin{equation}
  \int_0^1 \d x \, \frac{f(x)}{(1-x)_+}
  \equiv
  \int_0^1 \d x \, \frac{[f(x)-f(1)]}{(1-x)}.
\end{equation}
A simplification is obtained by observing that, under the integral, the plus
regularisation obeys the following identity:
\begin{eqnarray}
  \int_x^1 && \d y \, \left( \frac{y^n}{1-y} \right)_+ \, f(y)
\OneTwoCol{}{\nonumber\\&&}
  \equiv
  \int_x^1 \d y \, \frac{y^n}{(1-y)_+} \, f(y)
  + \sum_{m=1}^n \frac1m \, f(1).
\end{eqnarray}

The integral to evaluate may therefore be cast into the form
\begin{eqnarray}
  \int_x^1
  && \frac{\d y}{y}
  \left[ y \left( \frac{1+y^2}{1-y} \right)_+ \right]
  f\argleft( \frac{x}{y} \right)
\nonumber\\
  && \equiv
  \int_x^1 \d y \,
  \left( \frac{1+y^2}{1-y} \right) \,
  f\argleft( \frac{x}{y} \right)
  - \int_0^1 \d y \,   \left( \frac{1+y^2}{1-y} \right) \, f(x)
\nonumber\\
  && =
  \int_x^1 \d y \,
  \left( \frac{1+y^2}{1-y} \right) \,
  \left[ f\argleft( \frac{x}{y} \right) - f(x) \right]
\OneTwoCol{}{\nonumber\\&&\qquad\null}
  -
  \int_0^x \d y \, \left( \frac{1+y^2}{1-y} \right) \, f(x).
\end{eqnarray}
We choose to compute the integrals using the $x$ variable rather than the
transformed variable $u$ introduced in the main text for simplicity of
calculation; note that since the $x$-step ratio is always rather close to
unity, there is negligible numerical difference between the two approaches.

The integrals may be simplified via integration by parts to obtain
\begin{eqnarray}
  &&
  -\left[
    \left( 2\ln(1-y) + y + \frac12y^2 \right)
    \left( f\argleft(\frac{x}{y}\right) - f(x) \right)
   \right]_x^1
\nonumber\\
  && \quad \null
  +\int_x^1 \d y \,
   \left( 2\ln(1-y) + y + \frac12y^2 \right) \,
   \frac{\d}{\d y} f\argleft(\frac{x}{y}\right)
\OneTwoCol{}{\nonumber\\&&\qquad\null}
  +\left[
    2\ln(1-y) + y + \frac12y^2
   \right]_0^x \, f(x),
\end{eqnarray}
in which careful examination reveals that the first and third terms completely
cancel. Upon substitution of $y\to{x}/{y}$ we finally obtain
\begin{equation}
  -\int_x^1 \frac{\d y}{y} \, \frac{x}{y} \,
  \left[
    \ln \argleft( 1 - \frac{x}{y} \right) + \frac{x}{y} + \frac12\frac{x^2}{y^2}
  \right] \,
  \frac{\d}{\d y} f(y).
\end{equation}

The integral to be performed requires the division of the interval
$[\xmin,1]$ into $n$ sub-intervals of equal size in the variable $u$,
introduced earlier. While one might directly apply a trapezoidal (or better)
rule, since the integrand is a product of the unknown distribution function
and a known splitting function (with singularities), it is clearly better to
use the latter as a weight function. Thus, if we define the $k$-th interval as
$[x_k,x_{k-1}]$, with $x_k\equiv\lambda^k$ (and $\xmin=\lambda^n$), then from
eq.~(\ref{eq:evol-subs}) we have
\begin{eqnarray}
  &-&\sum_{k=1}^m \int_{x_k}^{x_{k-1}} \frac{\d y}{y} \, \frac{x_m}{y}
  \left[
    \ln \argleft( 1 - \frac{x_m}{y} \right)
    + \frac{x_m}{y}
    + \frac12\frac{x_m^2}{y^2}
  \right]
\OneTwoCol{}{\nonumber\\&&\hspace{8em}\null\times}
  \frac{(f_k - f_{k-1})}{(x_k - x_{k-1})},
  \hspace*{2em}
\end{eqnarray}
where we have identified $x$ with $x_m$ and assumed for simplicity a two-point
interpolation formula for $f(y)$. The coefficients of $f_k$ for each $x_m$
are then just the elements $P_{mk}$ of the DGLAP kernel matrix required.

It is immediately clear that precision will be poorer in the region $x\to1$.
However, two points should be borne in mind: firstly, the data are sparse in
this region and thus high precision is superfluous, and secondly, the absolute
value of $q(x)$ there is so small as to be negligible. Moreover, as and when
greater precision for $x\to1$ might be necessary, at no more than the cost of
doubling computer time, evolution could be performed in parallel for the
entire region of interest and for just the region $x\in[0.2,1]$, say.

In concluding this appendix let us remark that the form in which the elements
are calculated numerically is crucial for the avoidance of large round-off
errors due to repeated division and multiplication by $1-\lambda$, where
$\lambda$ is defined just after eq.~(\ref{eq:evol-sol}). We note also in
passing that as the interval size tends to zero, the coefficients diverge only
logarithmically, as all pole terms explicitly cancel. We further note that
the use of higher-order interpolation formul{\ae} leads to significantly greater
precision: we have tested Lagrange polynomials up to order seven and found
improvements in precision of rather more than two orders of magnitude.
\section{Proof of Banded Form of the DGLAP Matrices}
\label{app:proof}

Here we prove the statement that the matrix elements $P_{km}$ only depends on
$k-m$. First of all we note that the plus regularisation involves the
subtraction of an integral over the entire interval $[0,1]$, which therefore
depends neither on $m$ nor $k$, indeed, it contains no $x$ dependence, except
in the factor $f(x)$ itself. Thus, for the proof we may ignore the subtleties
of the plus regularisation, then the typical integral to evaluate takes the
form
\begin{equation}
  \sum_{k=1}^m \int_{x_k}^{x_{k-1}} \frac{\d y}{y} \,
  P\argleft(\frac{x_m}{y}\right) \, f(y).
\end{equation}
The two-point interpolating (Lagrange) function for $f(y)$ between the points
$x=x_k$ and $x_{k-1}$ is just
\begin{equation}
  f(y) \approx \frac{y - x_k    }{x_{k-1} - x_k    } \, f_{k-1}
             + \frac{y - x_{k-1}}{x_k     - x_{k-1}} \, f_k
\end{equation}
Making the substitution $y=x_{k-1}\xi$ and using $x_k=\lambda^k$ we obtain
\begin{equation}
  \sum_{k=1}^m \int_1^\lambda \frac{\d\xi}{\xi} \,
  P\argleft(\frac{\lambda^{m-k+1}}{\xi}\right) \, f(\lambda^{k-1}\xi),
  \label{eq:scaled}
\end{equation}
where the interpolation formula now becomes
\begin{equation}
  f(\lambda^{k-1}\xi)
  \approx
    \frac{\xi-\lambda}{1-\lambda} \, f_{k-1}
  + \frac{\xi-1      }{\lambda-1} \, f_k.
\end{equation}
Thus the final result for the coefficients of the $f_k$, in the integral
governing the evolution at $x_m$, clearly only depend on the difference
$m-k$, which in turn only appears in the numerator of the argument of $P$ in
eq.~(\ref{eq:scaled}). Clearly, increasing the number of interpolation points
does not affect the argument, \emph{provided the set of points used for any
given bin terminates at the lower $x$ boundary of that bin}.

\end{document}